\documentclass[aps,pra,twocolumn,amsmath,amssymb,showpacs,floatfix,superscriptaddress,nofootinbib]{revtex4-2}

\usepackage{times}
\usepackage{mdframed}

\usepackage{xcolor}
\usepackage[utf8x]{inputenc}
\usepackage[english]{babel}
\usepackage[T1]{fontenc}
\usepackage{lmodern}
\usepackage{amssymb}
\usepackage{amsthm}
\usepackage{amsmath}
\usepackage[pdftex]{graphicx}
\usepackage{epstopdf}
\usepackage{braket}
\usepackage[bottom]{footmisc}
\usepackage{bbold}
\usepackage{wasysym}
\usepackage{dcolumn}
\usepackage{bm}
\usepackage{color}
\usepackage{relsize,dsfont,mathrsfs,empheq,verbatim,upgreek}
\usepackage{etoolbox}
\usepackage[caption = false]{subfig}

\newtheorem*{myprop}{Proposition}

\newcommand{\Tr}{\mathrm{Tr}}
\newcommand{\Lt}{\mathcal{L}}
\newcommand{\Ht}{\mathcal{H}}

\newcommand{\hilbert}{\mathscr{H}}

\def\one{{\mbox{$1 \hspace{-1.0mm}  {\bf l}$}}}

\newcommand{\Tau}[2]{\boldsymbol{\tau}_{#1}^{#2}}
\newcommand{\Svec}[2]{\boldsymbol{s}_{#1}^{#2}}

\newcommand{\DcalR}[4]{\mathcal{D}_{\mathcal{R}}\!\left(\Tau{#1}{#2}\!, \Svec{#3}{#4}\right)}
\newcommand{\DcalI}[4]{\mathcal{D}_{\mathcal{I}}\!\left(\Tau{#1}{#2}\!, \Svec{#3}{#4}\right)}

\newcommand{\Xcal}[5]{{#1}\!\left(\Tau{#2}{#3}\!, \Svec{#4}{#5}\right)}

\begin{document}


\title{Unveiling coherent dynamics in non-Markovian open quantum systems: exact expression and recursive perturbation expansion }

\author{Alessandra Colla}
\email{alessandra.colla@unimi.it} 
\affiliation{Dipartimento di Fisica ``Aldo Pontremoli'', Universit\`a degli Studi di Milano, Via Celoria 16, I-20133 Milan, Italy}
\affiliation{INFN, Sezione di Milano, Via Celoria 16, I-20133 Milan, Italy}
\affiliation{Institute of Physics, University of Freiburg, 
Hermann-Herder-Stra{\ss}e 3, D-79104 Freiburg, Germany}

\author{Heinz-Peter Breuer}

\affiliation{Institute of Physics, University of Freiburg, 
Hermann-Herder-Stra{\ss}e 3, D-79104 Freiburg, Germany}

\affiliation{EUCOR Centre for Quantum Science and Quantum Computing,
University of Freiburg, Hermann-Herder-Stra{\ss}e 3, D-79104 Freiburg, Germany}

\author{Giulio Gasbarri}
\email{giulio.gasbarri@uni-siegen.de} 
\affiliation{Naturwissenschaftlich-Technische Fakult\"{a}t, Universit\"{a}t Siegen, 
Siegen 57068, Germany}
\affiliation{F\'isica Te\`orica: Informaci\'o i Fen\`omens Qu\`antics, Department de F\'isica, Universitat Aut\`onoma de Barcelona, 08193 Bellaterra (Barcelona), Spain}

\begin{abstract}
 We introduce a systematic framework to derive the effective Hamiltonian governing the coherent dynamics of non-Markovian open quantum systems. By applying the minimal dissipation principle, we uniquely isolate the coherent contribution to the time-local generator of the reduced dynamics. 
 We derive a general expression for the effective Hamiltonian and develop a recursive perturbative expansion that expresses it in terms of system-bath interaction terms and bath correlation functions. 
 This expansion provides a systematic tool for analyzing energy renormalization effects across different coupling regimes. 
 Applying our framework to paradigmatic spin systems, we reveal how environmental correlations influence energy shifts and eigenbasis rotations, offering new insights into strong-coupling effects and non-Markovian quantum thermodynamics.
\end{abstract}

\maketitle

\paragraph*{Introduction ---}
The dynamics of open quantum systems arise from the interplay between coherent evolution and environmental interactions~\cite{Breuer2007,Vacchini2024}. 
Beyond inducing dissipation and decoherence, these interactions can renormalize the energy structure of the system, giving rise to phenomena such as the Lamb~\cite{Bethe1947,Lamb1947} and AC Stark shifts~\cite{Aulter1955}. 
The Lamb shift, a prototypical renormalization effect caused by electromagnetic vacuum fluctuations, exemplifies how environmental couplings modify intrinsic system properties, with far-reaching implications for both theory and applications. 

Within the Lindblad framework, such shifts emerge naturally from tracing out environmental degrees of freedom under weak-coupling and Markovian assumptions~\cite{Breuer2007}. Extending this understanding to strong-coupling and non-Markovian regimes, however, remains an open challenge~\cite{Hossein-Nejad2015,Malavazi2022,Malavazi2022th,Nicacio2023,Neves2023,Neves2024,Tiwari2025}.

A powerful approach to non-Markovian dynamics is provided by time-convolutionless (TCL) master equations~\cite{Breuer2001,Breuer2007}, which describe exact reduced dynamics through a time-local generator $\mathcal{L}_{t}$. 
The generator can be written in a generalized Lindblad form~\cite{Hall2014,Breuer2012}:
\begin{align}\label{eq:no_gauge}
  \mathcal{L}_t[X] = -i [K(t), X] + \mathcal{D}_t[X],
\end{align}
where $K(t)$ denotes the effective Hamiltonian encoding coherent energy shifts, and $\mathcal{D}_t[X]$ represents the dissipative part, which may include negative rates ---a possible signature of non-Markovian behavior~\cite{Rivas_2014,Breuer2016C}. 

However, the decomposition of $\mathcal{L}_t$ into its coherent and dissipative components is not unique due to the intrinsic symmetry of the generator~\cite{Breuer2007,Chruscinski2022}. 
Resolving this ambiguity is essential for a consistent quantification of coherent energy shifts beyond the Markovian regime.

Hayden and Sorce recently proposed a foundational solution to this issue~\cite{Sorce2022}, defining the effective Hamiltonian by minimizing the dissipative contribution. 
This ``minimal dissipation'' approach isolates the coherent component of the dynamics and yields a dissipator with traceless Lindblad operators, recovering the standard weak-coupling limit where Hamiltonian renormalization is well understood. 
The physical relevance of this effective Hamiltonian has been experimentally confirmed in trapped-ion platforms operating in a highly non-Markovian regime~\cite{Colla2024exp}, where the resulting level shifts are typically time-dependent.

Beyond its foundational importance, the effective Hamiltonian also provides a key conceptual link to quantum thermodynamics in the strong-coupling and non-Markovian domains~\cite{Colla2022}. 
In quantum thermal machines, it captures energy renormalization effects arising from system–bath interactions, which crucially influence performance metrics such as efficiency and work extraction~\cite{Picatoste2024}. 
For time-dependent effective renormalization, $K(t)$ can describe work exchange between system and bath, revealing novel thermodynamic behaviors and challenging conventional energy-flow paradigms~\cite{Colla2022,Huang2022,colla2024FA}.

Despite its significance, determining $K(t)$ is often analytically or numerically demanding, especially in strongly coupled or high-dimensional systems, as time-local generators are rarely accessible in closed form. 
Exact expressions exist only for specific integrable models such as the Jaynes--Cummings~\cite{Smirne2010}, Fano--Anderson~\cite{Huang2022,Picatoste2024,colla2024FA}, and Caldeira--Leggett~\cite{Hu1992} Hamiltonians, linearly coupled Gaussian systems~\cite{Ferialdi2016}, and pure dephasing models~\cite{Luczka1990,Doll2008}. 
In general, $\mathcal{L}_t$ may be obtained numerically using methods such as hierarchical equations of motion (HEOM)~\cite{Tanimura1989} or pseudomode approaches~\cite{Tamascelli2018,Mascherpa2020}, or experimentally by quantum process tomography. 
In perturbative regimes, TCL expansions in the coupling strength provide an effective alternative.

In this Letter, we derive an operational expression for the effective Hamiltonian $K(t)$ directly from the time-local generator $\mathcal{L}_t$, enabling straightforward extraction of the coherent part of the dynamics once $\mathcal{L}_t$ is known. 
We further present a recursive perturbative expansion for $K(t)$ in powers of the system–bath coupling, expressed in terms of interaction operators and bath correlation functions. 
This construction allows systematic analysis of Hamiltonian renormalization across coupling regimes. 
We apply the method to paradigmatic spin models, illustrating how environmental characteristics affect the structure of the renormalized Hamiltonian.

\paragraph*{Effective Hamiltonian --}
The procedure recently proposed by Hayden and Sorce to uniquely fix the effective Hamiltonian is based on projecting the generator $\mathcal{L}$ into the subspace of superoperators that act as commutators with Hermitian operators, i.e. $\mathcal{H} = -i[H, \cdot]$, where $H = H^\dagger$~\cite{Sorce2022}. This projection is performed using a scalar product defined by:

\begin{align}\label{eq:sp_Haar}
\langle \mathcal{L}_1, \mathcal{L}_2 \rangle := \overline{\bra{\psi} \overline{\mathcal{L}_1[\ket{\phi}\bra{\phi}] \mathcal{L}_2[\ket{\phi}\bra{\phi}] } \ket{\psi}},
\end{align}
where the overline denotes averaging over Haar-random pure states $\ket{\phi}$ and $\ket{\psi}$.
This approach minimizes the dissipator assuming maximum uncertainty over input and output states (i.e. uniformly sampling over pure states), yielding a canonical effective Hamiltonian $K$ for any linear, Hermiticity-preserving, trace-annihilating generator $\Lt \in \mathfrak{htp}(\hilbert)$ (where $\hilbert$ denotes the Hilbert space of the reduced system).

The physical interpretation of this procedure is better understood via the average fidelity. 
The fidelity between a noiseless von Neumann generator $\mathcal{H} = -i[H, \cdot]$, associated with some Hamiltonian $H$, and the generator of a noisy channel described by $\mathcal{L}$ is defined as:
\begin{align}
  \overline{\mathcal{F}}(\mathcal{H}, \mathcal{L}) = \int d\psi\, \mathrm{Tr}\left\{\mathcal{H}[\ket{\psi}\bra{\psi}]\,\mathcal{L}[\ket{\psi}\bra{\psi}]\right\}.
  \end{align}
This quantity measures how ``faithfully'' the noisy dynamics generated by $\mathcal{L}$ approximates that of the noiseless $\mathcal{H}$ when applied to randomly chosen pure states, and is connected to the scalar product \eqref{eq:sp_Haar} between the generators simply through a scaling with the Hilbert space dimensions $d = \dim(\hilbert)$, i.e.
$    \braket{\mathcal{H},\mathcal{L}} = \frac{1}{d}\overline{\mathcal{F}}(\mathcal{H}, \mathcal{L})$.

Expanding $H$ in a basis of traceless Hermitian operators $\{H_j\}_j$, orthonormal under the Hilbert-Schmidt product, the effective Hamiltonian is expressed as (see Appendix~\ref{app:fidelity}):
\begin{align}\label{eq:Ks-fidelity}
K = \frac{d+1}{2} \sum_j \overline{\mathcal{F}}(\mathcal{H}_j, \mathcal{L}) H_j,
\end{align}
where $\mathcal{H}_{j} =-i [H_{j},\cdot]$
This expression for $K$ establishes a direct connection between the average fidelity and the effective Hamiltonian.
It reveals that the effective Hamiltonian is a weighted sum of basis operators $H_j$, with weights determined by the average fidelity between $\mathcal{L}$ and the generators corresponding to each basis operator.
It quantifies how well different components of the dynamics preserve quantum information on average and clarifies the equal apriori probability -- or maximum uncertainty -- principle embedded in the choice of the norm induced by the scalar product \eqref{eq:sp_Haar}.

Equation~\eqref{eq:Ks-fidelity} provides physical insight into $K$, but it is not practical for calculations.
Therefore, we provide a general expression for the effective Hamiltonian associated with the canonical splitting under
\eqref{eq:sp_Haar} of any given generator $\Lt \in \mathfrak{htp}(\hilbert)$, which constitutes our first result.
\begin{myprop}
  Let $\{F_{\alpha}\}_{\alpha=0}^{d^2-1}$ be any orthonormal basis of $\mathcal{B}(\hilbert)$ with respect to the Hilbert-Schmidt scalar product,
  \begin{align}\label{Ks_Fa}
  \Tr\{F_\alpha^{\dag}F_{\beta}\} = \delta_{\alpha \beta} \;, \forall\; \alpha , \beta \; .
  \end{align}
  Then, for any generator $\Lt \in \mathfrak{htp}(\hilbert)$ its corresponding effective Hamiltonian is
  \begin{align}\label{eq:K_general}
  K = \frac{1}{2i d} \sum_{\alpha=0}^{d^2-1} [F_{\alpha}^{\dag}, \mathcal{L}[F_{\alpha}]] \; .
  \end{align}
  \end{myprop}
 
The above equation can be validated by substituting $\Lt$ with a generic canonical generator satisfying minimal dissipation, i.e eq.~\eqref{eq:no_gauge} with traceless Lindblad operators. 
We further provide a detailed derivation of eq.~\eqref{eq:K_general} starting from the definition of the scalar product in  eq.~\eqref{eq:sp_Haar} 
(see Appendix~\ref{app:eq6}).
Using the operator basis $F_\alpha = |j\rangle\langle k|$, where $|j\rangle$ form an orthonormal basis on $\hilbert$, this becomes:
 \begin{align}\label{eq:K-jk}
 K = \frac{1}{2i d} \sum_{j,k=1}^d [|j\rangle\langle k|,\,\mathcal{L}[|k\rangle\langle j|]].
 \end{align}
These new expressions are applicable to a wide range of quantum systems and coupling regimes, and offer a more practical approach for computing $K$ whenever the exact generator is explicitly known. However, in most cases, an explicit form of $\Lt$ is only accessible via perturbative treatment. In these cases, the form of the generator is typically given by $\mathcal{L}[X] = \sum_{ij} \omega_{ij} V_{i} X W_{j}$, with $V_{i}$ and $W_{j}$ satisfying Hermiticity-preservation and trace-annihilation conditions. This leads to an effective Hamiltonian of the following form:
 \begin{align}\label{eq:K_rec}
 K = \frac{1}{2i d} \sum_{ij} \omega_{ij} \left( \mathrm{Tr}(V_i) W_j - \mathrm{Tr}(W_j) V_i \right),
 \end{align}
where the result from~\cite{Sorce2022} is recovered when $\omega_{ij}$ is diagonal (pseudo-Kraus representation). In what follows, we provide a technique to recover the contributions to the effective Hamiltonian at each order of the perturbation expansion.

\paragraph*{Systematic perturbation expansion --- }
The generator $\mathcal{L}_t$ of an open quantum system typically arises from the microscopic unitary evolution of a closed bipartite system, where the system $S$ interacts with an environment $E$ (the irrelevant degrees of freedom that are traced out). The total Hamiltonian governing this evolution is given by: 
\begin{align}
  H = H_S + H_E + \lambda H_I,
\end{align}
where $H_S$ and $H_E$ are the free Hamiltonians of the system and environment, respectively, and $\lambda$ is the coupling parameter for the interaction $H_I$. For simplicity, we consider a factorized interaction Hamiltonian, $H_I = A \otimes B$, where $A$ and $B$ act on the system and environment, respectively. Extensions to sums of such terms are straightforward.
We further assume the total initial state to factorize as $\rho_{SE}(0) = \rho_S \otimes \rho_E$, 
since initial system–environment correlations do not affect the coherent part of the system dynamics, i.e., the effective Hamiltonian $K$~\cite{Colla2022corr}.
Their only effect is the addition of an inhomogeneous term to the homogeneous map describing the evolution of the initially uncorrelated state, while the homogeneous part of the dynamics characterized by the generator in eq.~\eqref{eq:no_gauge} remains unchanged. See Appendix~\ref{app:correlations} for a detailed discussion.

Finding the exact effective Hamiltonian $K$ for the reduced system $S$ is generally challenging without an explicit expression for the time local generator $\mathcal{L}_{t}$ associated to the reduced dynamics of the system only. Here, we provide a systematic perturbation expansion for $K$ in powers of the coupling strength $\lambda$. 
This constitutes the second main result of this work. 
Below, we outline the strategy; technical details are provided in~\cite{Colla2025pra}. 

The interaction picture is used to simplify calculations, where the interaction Hamiltonian becomes time-dependent: $\tilde{H}_t = A(t) \otimes B(t)$.
The total unitary evolution can then be written as: 
\begin{align}
  \mathcal{U}_{t}[\rho_{SE}(0)] = \mathcal{T}\left( e^{-i \int_0^t d\tau (\tilde{H}\tau^L - \tilde{H}\tau^R)} \right) [\rho_{SE}(0)],
\end{align} 
where $\mathcal{T}$ denotes the time ordering and $\tilde{H}_\tau^{L/R}$ represents the operator acting on $\rho$ from the left and from the right~\cite{Diosi1993,Diosi1990,Chou1984es}.
 
The reduced dynamics of the system is obtained by 
tracing out the environment degrees of freedom, yielding to  
 $  \Phi_t[\rho_S] = \mathrm{Tr}_E \{\mathcal{U}_t[\rho_S \otimes \rho_E]\} $.
  Expanding the unitary evolution in a Taylor series, we express $\Phi_{t}$ as: 
  
  \begin{align}
  \Phi_t = 1 + \sum_{n=1}^\infty (-i)^n \lambda^n \mu_n,
  \end{align} 
where each term $\mu_n$ depends on time-ordered integrals over system and bath operators. These contributions can be decomposed as: 
\begin{align}
  \mu_n^k & = \int_0^t d\boldsymbol{\tau}_1^k d\boldsymbol{s}_1^{n-k} D(\boldsymbol{\tau}_1^k, \boldsymbol{s}_1^{n-k}) A^L(\boldsymbol{\tau}_1^k) A^R(\boldsymbol{s}_1^{n-k}),
  \end{align}
where we have defined
\begin{align}
{A^{L/R}}(\boldsymbol{t}_1^k)[\,\cdot\,] &:= A^{L/R}_{t_1}\circ{A^{L/R}}_{t_2}\circ...\circ{A^{L/R}}_{t_k} [\,\cdot\,] \; ,
\end{align}
and with shorthand integral notation $\int_{0}^{t}d\boldsymbol{\tau}_{1}^{k} d\boldsymbol{s}_{1}^{n-k}$ indicating integration over $k$ times ${\tau_1, ..., \tau_k}$ and $(n-k)$ times ${s_1, ..., s_{n-k}}$. Here, $D(\boldsymbol{\tau}_1^k, \boldsymbol{s}_1^{n-k})$ encodes environmental correlations via $n$-point bath correlation functions:
    \begin{align}
  D(\boldsymbol{\tau}_1^k, \boldsymbol{s}_1^{n-k}) = & \mathrm{Tr} \left\{B^R(\boldsymbol{s}_1^{n-k}) B^L(\boldsymbol{\tau}_1^k) \rho_E \right\}
  \nonumber\
   \theta_{\boldsymbol{\tau}_1^k} \theta_{\boldsymbol{s}_1^{n-k}},
  \end{align} 
  where $\theta_{\boldsymbol{\tau}_{1}^{n}}$ ensures time ordering within each set of variables $\{\tau_{j}\}_{j\in[1,n]}$, i.e. $\theta_{\boldsymbol{\tau}_{1}^{n}} $ is equal to 1 if  $t_{1}> t_{2}>\dots\tau_{n}$ and zero otherwise.
  
Following the steps in~\cite{Colla2025pra,Gasbarri2018} the generator can be expressed as: 
\begin{align}
    \mathcal{L}_{t} = \sum_{n=0}^\infty \lambda^n \mathcal{L}_n,
\end{align} 
where each term $\mathcal{L}_n$ depends on combinations of $\mu_n$ terms and their time derivatives. (For a detailed derivation the perturbative series see~\cite{Colla2025pra}) 
Applying eq.~\eqref{eq:K_rec} for each order of $\mathcal{L}_n$, we obtain a perturbative expansion for the effective Hamiltonian: 
\begin{align}
  K(t) = \sum_{n=0}^{\infty} \lambda^n K_n.
\end{align} 
With notation $\braket{X}_{1/d}:= \Tr\{X\}/d$, and defining the operator $\Xcal{\mathbb{A}}{1}{k}{1}{n-k} := \langle A(\boldsymbol{s}_{1}^{n-k})^{\dagger}\rangle_{1/d}  A(\boldsymbol{\tau}_{1}^{k})$, the $n$th-order contribution to $K(t)$ is given by:
 \begin{widetext}
\begin{align}\label{eq:Kn-final}
 K_n
&= \frac{-(+i)^n}{2i}\sum_{k=0}^n (-)^{k} \int_0^t d \boldsymbol{\tau}_{1}^{k} d \boldsymbol{s}_{1}^{n-k} \left[ \mathcal{D}(\boldsymbol{\tau}_{1}^{k},\boldsymbol{s}_{1}^{n-k})\Xcal{\mathbb{A}}{1}{k}{1}{n-k} - (-)^{n} \text{h.c.}\right] \; ,
\end{align}
\end{widetext}
where the functions $\mathcal{D}$ are bath $n$-ordered cumulants and are linear combinations of products of the $n$-point correlation functions (see companion paper~\cite{Colla2025pra}) which can be recovered via the following recursion formula:
\begin{align}\label{eq:Dcal-rec}
  \mathcal{D}(\boldsymbol{\tau}_{1}^{k},\boldsymbol{s}_{1}^{n-k}) &=
  \dot{D}(\boldsymbol{\tau}_{1}^{k},\boldsymbol{s}_{1}^{n-k}) \nonumber\\
  & \quad -\sum_{l=0}^{k}\sum_{r=0}^{n-k}   \mathcal{D}(\boldsymbol{\tau}_{1}^{l},\boldsymbol{s}_{1}^{r}){D}(\boldsymbol{\tau}_{l+1}^{k},\boldsymbol{s}_{r+1}^{n-k})
\end{align}
where
\begin{align}
\dot{D}(\boldsymbol{\tau}_{1}^{k},\boldsymbol{s}_{1}^{n-k})=D(\boldsymbol{\tau}_{1}^{k},\boldsymbol{s}_{1}^{n-k})
 \big( \delta_{\tau_1,t}+\delta_{s_{1},t} \big) \; ,
\end{align}
and we have taken the convention $\int_0^td\tau \delta_{\tau, t} f(\tau) = f(t)$ for any function of $f$. The time ordering functions $\theta$ inside each term $D$ are responsible for the time ordering of all operators and functions in the integrals, and will give rise to a number of different terms that are ordered in ``chunks'', similarly to the ordered cumulants of van Kampen~\cite{VanKampen1974a,VanKampen1974b}.

These expressions allow systematic computation of $K(t)$ at any order in $\lambda$, revealing how system-bath interactions renormalize energy levels. Notably, eq.~\eqref{eq:Kn-final} provides explicit information about the structure of $K(t)$ in terms of system operators ($A$) and their interaction-picture dynamics and the $n$-point correlation functions of the bath.

\paragraph*{Spin systems --- }
To elucidate the structure of the effective Hamiltonian, we apply our results to open spin systems. Without loss of generality, we take $H_S = \frac{\omega}{2}\sigma_z $.
The recursive series for $K$ here derived reveals whether the interaction with a thermal bath generates only a shift of the energy levels of the free Hamiltonian or also a rotation of its eigenvectors. 

Suppose that $A=\sigma_z$, resulting in a situation of pure dephasing for the spin, where $A(t)\equiv A$ $\forall t$.
Using \eqref{eq:Kn-final}, we find (see Appendix~\ref{app:spin}) that the contribution of the odd terms of the series commute with the free Hamiltonian, and the even terms are vanishing.
This leads to an effective Hamiltonian for pure dephasing of the form
\begin{align}\label{ex:K_s_diag}
K(t) = \frac{\omega_r(t)}{2}\sigma_z \;,
\end{align}
where only the odd ordered cumulants of the bath contribute to the time-dependent energy shift.

This result agrees with time-dependent shifts obtained in~\cite{Seegebrecht2024} (dispersive spin oscillator) and in~\cite{Neves2024} (pure qubit-qubit dephasing).
It furthermore explains the absence of renormalization found in~\cite{Seegebrecht2024} (displaced spin oscillator), and in the well-known dephasing spin-boson model~\cite{leggett1987dynamics}:
\begin{align}
H = \frac{\omega}{2}\sigma_z + \sum_j \omega_j a^\dag_j a_j + \sum_j g_j \sigma_z \otimes (a^\dag_j + a_j)
\end{align}
with an initial state of the environment that is diagonal in the Fock basis (including, in particular, a thermal state), for which the exact TCL master equation (already exact at second order~\cite{Doll2008}) is well known to be
\begin{align}
\dot{\rho}_S(t) = -i \left[\frac{\omega}{2}\sigma_z, \rho_S(t)\right] + \gamma(t) [\sigma_z \rho_S(t)\sigma_z - \rho_S(t)]\;,
\end{align}
where $\gamma(t)$ is the decoherence rate.

This drastic difference between the nature of even and odd contributions of the series is not just limited to dephasing models. Let us consider an unbiased spin model, where the interaction term is described by an operator that is orthogonal, under the Hilbert-Schmidt scalar product, to the Hamiltonian term, such as $A=\sigma_x$.

In this case, the odd terms of the series do not commute with the free Hamiltonian, i.e., $[K_{2m+1},H_S]\neq 0$, inducing a rotation of the eigenvectors of the Hamiltonian (see Appendix~\ref{app:spin}). 
In contrast, the even terms do commute with $H_S$, contributing to a shift in the energy spectrum. Thus, when odd orders vanish, the effective Hamiltonian only results in a renormalization of the spin frequency, as expressed in eq.~(\ref{ex:K_s_diag}).
Notably, this is the case for a spin-boson model at finite temperature: 
 \begin{align}
   H = \frac{\omega}{2} \sigma_z + \sum_j \omega_j a^\dag_j a_j + \sum_j g_j \sigma_x \otimes (a^\dag_j + a_j) \;.
  \end{align} 
This result proves analytically what was conjectured in~\cite{Gatto2024} due to numerical evidence.

Commonly studied scenarios are those in which the dynamics is characterized by two-point correlation functions, as in the case of linearly coupled, zero-mean Gaussian baths. 
The present analysis clarifies why such models exhibit simpler renormalization effects: in the case of pure dephasing, renormalization is entirely absent, while for orthogonal couplings, the renormalized Hamiltonian remains diagonal in the original eigenbasis. In contrast, when the bath distribution exhibits time asymmetry (i.e., when odd-order cumulants are nonzero), more intricate behaviors emerge, such as rotations of the Hamiltonian eigenbasis.

\paragraph*{Structure of the Hamiltonian ---}

The examples above show that the effective Hamiltonian exhibits qualitatively distinct behaviors depending on the interaction type and environmental structure. 
More generally, the perturbative expansion reveals a fundamental symmetry difference between even and odd orders. This feature is not limited to spin models: it arise from decomposing the ordered cumulants of the bath into real and imaginary parts, $\mathcal{D}=\mathcal{D}_{\mathcal{R}}+i\mathcal{D}_{\mathcal{I}}$, which are respectively symmetric and antisymmetric under time ordering reversal. A similar decomposition holds for the system contribution, where $\Xcal{\mathbb{A}}{1}{k}{1}{n-k}$ splits into $\Xcal{\mathbb{A}_{\mathcal{R}}}{1}{k}{1}{n-k}$ and $\Xcal{\mathbb{A}_{\mathcal{I}}}{1}{k}{1}{n-k}$, which respectively are symmetric and antisymmetric under time ordering reversal. 

Making these decompositions explicit in eq.~\eqref{eq:Kn-final} we find that even-order terms $K_{2m}$ couple bath and system contributions of opposing time symmetry classes, namely
\begin{align}\label{eq:Kn-even-fd}
    K_{2m}\! =\!\!\! \sum_{k=0}^{2m} &
 (-)^{m+k+1} \!\!\int_0^t\!\!\!\! d \boldsymbol{\tau}_{1}^{k} d \boldsymbol{s}_{1}^{n-k}\!\! \left[ \DcalR{1}{k}{1}{n-k}\Xcal{\mathbb{A}_\mathcal{I}}{1}{k}{1}{n-k}\right.\nonumber\\
 &\left.+ \DcalI{1}{k}{1}{n-k}\Xcal{\mathbb{A}_\mathcal{R}}{1}{k}{1}{n-k}\right] \; ,
 \end{align}
ensuring an overall antisymmetric structure under time-ordering reversal.
Conversely, the odd terms $K_{2m+1}$ couple terms of the same symmetry class and are overall symmetric 
\begin{align} 
    \label{eq:Kn-odd-fd}
    K_{2m\!+\!1} \!=\!\!\!\! \sum_{k=0}^{2m+1}\!\!&
    (-)^{m+k}\!\!\! \int_0^t\!\!\!\!\! d \boldsymbol{\tau}_{1}^{k} d \boldsymbol{s}_{1}^{n-k}\!\! \left[ \DcalI{1}{k}{1}{n-k}\!\Xcal{\mathbb{A}_\mathcal{I}}{1}{k}{1}{n-k}\right.\nonumber\\
    &\left.- \DcalR{1}{k}{1}{n-k}\Xcal{\mathbb{A}_\mathcal{R}}{1}{k}{1}{n-k}\right] \;,
\end{align}
highlighting the fundamentally distinct symmetry properties and structure of even and odd orders in the effective Hamiltonian.

From this perspective, we can clearly interpret the terms up to second order.
The first order contribution arises only if the interaction-picture bath operator has a non-zero average with respect to the bath's initial state -- often implying the presence of non-zero initial coherences in the environment~\cite{colla2024FA} -- and corresponds to a driving of the system resulting from a displacement of the bath state relative to the interaction Hamiltonian. 

The second order contribution, instead, can be understood via linear response theory~\cite{Kubo1957} as arising from fluctuations and dissipation in the environment. In this framework, dissipation is associated with
the imaginary part of the correlation function $\chi(t, s) := i \langle [{B}(t),{B}(s)]\rangle$, i.e. the response function, which
is antisymmetric under time reversal. This component
reflects the causal response of the environment and accounts for the irreversible energy exchange associated with it. Diffusion, on the other hand, is associated
to the real part of the correlation function $S(t, s) := \frac{1}{2} \langle \{{B}(t),{B}(s)\}\rangle - \langle {B}(t) \rangle\langle {B}(s) \rangle$, which is symmetric under time reversal. This symmetry represents the environmental fluctuations that drive random perturbations in the system state. 
These random perturbations are responsible for
processes such as decoherence, and the spreading of the
system’s distribution over time.
In terms of these quantities, the second order contribution to $K$ can be written as
\begin{align}
    K_2 = & \int_0^t d\tau \Big[\frac{1}{2i}S(t,\tau) [A(t),A(\tau)] \\
    &-  \frac{\chi(t,\tau)}{4}\left(\{A(t),A(\tau)\} - \langle\{A(t),A(\tau)\}\rangle_{1/d}\right) \Big] \; . \nonumber
\end{align}

These insights suggest that distinct environmental features underpin the structural differences observed between even and odd orders of the renormalized Hamiltonian. Although our explicit analysis has been limited to second-order terms, it is reasonable to expect higher-order terms to similarly reflect more intricate environmental correlations.

\emph{Conclusions -- }
We introduced a systematic framework to obtain the effective Hamiltonian governing coherent dynamics in non-Markovian open quantum systems. By employing the minimal dissipation principle, we uniquely isolated the coherent component of the time-local generator. We provided an operational expression enabling the direct extraction of the effective Hamiltonian from the generator, as well as a recursive perturbative expansion in terms of bath correlation functions and system-bath interactions.

Our perturbative analysis clarified how environmental correlations and their interplay with system operators lead to energy renormalization, demonstrating clear structural distinctions between even and odd orders of the expansion. These distinctions are used to explain recently observed behaviors in spin systems, including eigenbasis rotations and time-dependent energy shifts.
Moreover, our perturbative series can serve as a complementary tool alongside established numerical approaches such as the HEOM, providing analytical guidance and a means to interpret numerical results. This synergy is illustrated in our spin system example, where the perturbative insights help corroborate and elucidate the behavior observed in simulations.
Together, these advances have direct relevance for strong-coupling quantum thermodynamics and the engineering of quantum thermal machines.

\acknowledgments
A.C. acknowledges support from MUR via the PRIN 2022 Project “Quantum Reservoir Computing (QuReCo)” (contract n. 2022FEXLYB). A.C. and H.P.B. acknowledge financial support from the European Union's Framework Programme for Research and Innovation Horizon 2020 (2014-2020) under the Marie Sk\l{}odowska-Curie Grant Agreement No.~847471.
G.G. acknowledges financial support by the DAAD, the Deutsche
Forschungsgemeinschaft (DFG, German Research Foundation, project numbers 447948357 and 440958198), the SinoGerman Center for Research Promotion (Project
M-0294), the German Ministry of Education and Research
(Project QuKuK, BMBF Grant No. 16KIS1618K), and the Stiftung Innovation in der Hochschullehre and the Spanish MICIN (project PID2022-141283NB-I00) with the support 
of FEDER funds.

\appendix

\section{$K$ in terms of generator fidelity}\label{app:fidelity}
The scalar product defined in~\cite{Sorce2022} for two generators reads
\begin{equation}\label{eq:sp_haar}
\braket{\mathcal{L}_1,\mathcal{L}_2} = \int\! d \phi \! \int\! d\psi \bra{\phi}\mathcal{L}_1\left[\ket{\psi}\!\!\bra{\psi}\right] \mathcal{L}_2 \left[\ket{\psi}\!\!\bra{\psi}\right] \ket{\phi} \;,
\end{equation}
where $\psi$ and $\phi$ are Haar-random generated pure states.
The average fidelity~\cite{Nielsen2002} between generators of a gate and a noisy channel defined in the main text is then directly proportional to the scalar product \eqref{eq:sp_haar} of the generators:
\begin{equation}\label{sp_fidelity}
\overline{\mathcal{F}}(\mathcal{H}, \mathcal{L})= \int d\psi \Tr\left\{\mathcal{H}\left[\ket{\psi}\bra{\psi}\right] \mathcal{L} \left[\ket{\psi}\bra{\psi}\right] \right\}  = d \braket{\mathcal{H},\mathcal{L}}\;,
\end{equation}
since $\int d \phi \bra{\phi} X \ket{\phi} = \Tr\{X\}/d$.
Then, in~\cite{Sorce2022} one chooses a basis of traceless Hamiltonians $\{H_j\}_j$ such that their associated generators $\mathcal{H}_j$ form an orthonormal basis for $\mathfrak{ham}(\hilbert)$ under \eqref{eq:sp_haar}, and find the effective Hamiltonian to be given by $K = \sum_j H_j \braket{\mathcal{H}_j,\Lt}$. 
The Hamiltonian elements $H_j$ are orthogonal under the Hilbert-Schmidt scalar product but not orthonormal, as $\Tr\{H_j^2\}= d(d+1)/2$. By defining a normalized basis for Hilbert-Schmidt 
\begin{eqnarray}
    \tilde{H}_j= \sqrt{\frac{2}{d(d+1)}}H_j
\end{eqnarray}
we obtain the equation for the effective Hamiltonian in terms of generator fidelity i.e. eq.~(4).

\section{Derivation of eq.~\eqref{eq:K_general}} \label{app:eq6}
The derivation is given as in~\cite{Sorce2022} and~\cite{Colla2022}  by projecting the generator onto the subspace $\mathfrak{ham}(\hilbert)$ of superoperators acting as $-i$ times a commutator with a traceless Hamiltonian,
\begin{equation}
\Ht = \Pi(\Lt) = \sum_j \Ht_j \langle \Ht_j , \Lt \rangle \; ,
\end{equation}
with $\{\Ht_j\}_{j=1}^{d^2-1}$ an orthonormal basis of $\mathfrak{ham}(\hilbert)$, through the scalar product \eqref{eq:sp_haar} on $\mathfrak{htp}(\hilbert)$. Given the Hamiltonians $H_j$ associated to the commutators in each $\Ht_{j}$, which obey $\Tr \{ H_i H_j\} = \frac{d(d+1)}{2} \delta_{ij}$, the effective Hamiltonian $K$ associated to $\Ht$ reads $K = \sum_j H_j \langle \Ht_j , \Lt \rangle $~\cite{Colla2022}.
Using the second moment Haar integral and the cyclic properties of the trace, the coefficients $\langle \Ht_j , \Lt \rangle$ are given simply by
\begin{equation}
\langle \Ht_j , \Lt \rangle = -\frac{i}{d} \Tr\{H_j\overline{\big[\ket{\phi}\bra{\phi}, \Lt [ \ket{\phi}\bra{\phi}]\big]}\} \; .
\end{equation}
Given that for any operator $X$ it holds that $\sum_j H_j \Tr \{ H_j X\}  = \frac{d (d+1)}{2 } X$, we find
\begin{align}
K =& \frac{d+1}{2i} \overline{\big[\ket{\phi}\bra{\phi}, \Lt [\ket{\phi}\bra{\phi}]\big]} \nonumber \\ =& (d+1) \text{Im}\Big(\overline{\ket{\phi}\bra{\phi}\Lt [\ket{\phi}\bra{\phi}]}\Big) \; .
\end{align}
To evaluate the fourth moment Haar integral for an unknown generator, consider an orthonormal basis $\{\ket{i}\}_{i=1}^{d}$ of $\hilbert$, and take a specific element $\ket{\bar{k}} =: \ket{\phi_0}$ as reference state such that $\ket{\phi} = U\ket{\bar{k}} = \sum_i u_{i\bar{k}} \ket{i}$ with $U$ a Haar random unitary transformation. Then
\begin{equation}
\ket{\phi}\bra{\phi}\Lt [\ket{\phi}\bra{\phi}] = \sum_{ijkl} u_{i\bar{k}}u_{k\bar{k}}u^*_{j\bar{k}}u^*_{l\bar{k}} \ket{i}\!\!\bra{j} \Lt[\ket{k}\!\!\bra{l}] \; .
\end{equation}
The Haar average therefore acts only on the unitary transformation elements, and the fourth moment is given in terms of the Weingarten functions $\text{Wg}$,~\cite{Gessner2013}:
\begin{align}
\langle u_{i\bar{k}}u_{k\bar{k}}u^*_{j\bar{k}}u^*_{l\bar{k}}\rangle =& (\delta_{ij}\delta_{kl}\delta_{\bar{k}\bar{k}} +\delta_{il}\delta_{jk}\delta_{\bar{k}\bar{k}})(\text{Wg}[1] + \text{Wg}[2])\nonumber \\ =& \frac{1}{d(d+1)} (\delta_{ij}\delta_{kl}+ \delta_{il}\delta_{jk}) \; ,
\end{align}
which gives
\begin{align}
\overline{\ket{\phi}\!\!\bra{\phi}\Lt [\ket{\phi}\!\!\bra{\phi}]} =  \frac{1}{d(d+1)} \sum_{jk}&\Large( \ket{j}\!\!\bra{j}  \Lt[\ket{k}\!\!\bra{k}]  \nonumber \\ &  +  \ket{j}\!\!\bra{k} \Lt[\ket{k}\!\!\bra{j}]  \Large)\; .
\end{align}
The first term is Hermitian and drops out of the imaginary part, so that
\begin{align}
K =& \frac{1}{d} \sum_{j,k=1}^d \text{Im}\Big( \ket{j}\bra{k}\mathcal{L}[\ket{k}\!\!\bra{j}]\Big) \nonumber \\ =& \frac{1}{2i d} \sum_{j,k=1}^d \big[\ket{j}\!\!\bra{k}, \mathcal{L}[\ket{k}\!\!\bra{j}]\big] \; . \label{eq:Ks_jk}
\end{align}
Since the choice of the basis $\ket{i}$ is arbitrary, one can perform any unitary transformation of the basis $\{\ket{j}\!\!\bra{k}\}_{j,k}$ onto any basis of operators $F_{\alpha}$ that is orthonormal with respect to the Hilbert-Schmidt product of operators, see e.g.~\cite{Gorini1976a}, to find  eq.~\eqref{eq:K_general}.

\subsection*{Basis choice: generators of $\mathfrak{su}(d)$}
Let us choose the orthonormal basis $F_0=\mathbb{1}/\sqrt{d}$ and $F_i=\sigma_i/\sqrt{d}$ for $i\neq 0$ with $\sigma_i$ generators of $\mathfrak{su}(d)$, with algebraic relations
\begin{equation}
\sigma_j \sigma_k = \delta_{jk} +\sum_{l=1}^{d^2-1} (i f_{jkl}+ d_{jkl})\sigma_l
\end{equation}
where $f_{jkl}$ are the completely antisymmetric structure constants of $\mathfrak{su}(d)$, and $d_{jkl}$ completely symmetric.
Then the contribution from $F_0$ drops out and the effective Hamiltonian is given in terms of the generators only:
\begin{equation}\label{Ks_sigma}
K = \frac{1}{2i d} \sum_{j=1}^{d^2-1} \left[\sigma_j, \mathcal{L}[\sigma_j] \right] \; .
\end{equation}
This expression might be particularly useful for systems with low-dimensional Hilbert spaces. If we also write the generator in terms of this basis, $\Lt=\sum_{\alpha \beta} l_{\alpha \beta} F_{\alpha} \cdot F_{\beta}$ we get simply
\begin{equation}
K = \frac{1}{2i d} \sum_{j}  \left(l_{0j} - l_{j0} \right) \sigma_j \; .
\end{equation}

\section{Initial correlations} \label{app:correlations}
Following~\cite{Colla2022corr}, initial system--environment correlations do not modify the Hamiltonian part of the reduced dynamics. The composite initial state can be written as
\begin{align}
\rho_{SE}(0) = \rho_S \otimes \rho_E + \chi,
\end{align}
 where $\rho_S = \Tr_E\{\rho_{SE}\}$, $\rho_E = \Tr_S\{\rho_{SE}\}$, and the correlation operator 
$\chi = \rho_{SE} - \rho_S \otimes \rho_E$ satisfies $\Tr_E\{\chi\} = \Tr_S\{\chi\} = 0$. 
The presence of $\chi$ introduces an inhomogeneous contribution to the reduced dynamics, leading to an affine time-local generator that can be uniquely extended to a linear map on the space of trace-one operators, without affecting the Hamiltonian term. Explicitly
\begin{align}\label{eq:corrgenerator}
\mathcal{L}^{\chi}_t [ X ] = \mathcal{L}_t[X] + \mathcal{J}^{\chi}_t \Tr \{ X\} \; ,
\end{align}
where $\mathcal{L}_t$ is the generator for initially uncorrelated states and
\begin{align}
\mathcal{J}_t^\chi = \dot{I}_t^\chi - \mathcal{L}_t[I_t^\chi] \; , \quad I_t^{\chi} = \Tr_E \{ U_t \chi U_t^{\dagger} \} \; .
\end{align}
The effective Hamiltonian in the presence of initial correlations can then be identified by applying eq.~\eqref{eq:K_general} to the generator in eq.~\eqref{eq:corrgenerator}. Importantly, this yields the same Hamiltonian part $K(t)$ as in the uncorrelated case. Thus, initial correlations affect only the dissipative (non-Hamiltonian) part of the evolution, leaving the  effective Hamiltonian invariant.

\section{Contributions to $K$ in spin systems}\label{app:spin}
\subsection{Pure decoherence}
Suppose that $A=\sigma_z$, so that $A(t)\equiv A$ $\forall t$, namely a situation of pure decoherence of the spin. Then we have $\Xcal{\mathbb{A}}{1}{k}{1}{n-k} = \langle \sigma_z^{n-k}\rangle_{1/d} \sigma_z^{k}$, which is easily found depending on the parity of $n$ and $k$:
\begin{equation}
\Xcal{\mathbb{A}}{1}{k}{1}{n-k} = \begin{cases}
    \sigma_z & \text{($n$ odd, $k$ odd)} \\
    \mathbb{1} & \text{($n$ even, $k$ even)}  \\
    0 & \text{elsewhere}
\end{cases}\; .
\end{equation}
This means that there are no even order terms in the effective Hamiltonian -- as the final Hamiltonian is always traceless and any term proportional to the identity will vanish -- and that any odd order term will give a contribution proportional to $\sigma_z$.
Thus, we conclude that for a spin under pure decoherence, the effective Hamiltonian is in general of the form $K(t) = \frac{\omega(t)}{2}\sigma_z$.

\subsection{Unbiased spin system}
Suppose the interaction term to be orthogonal to $\sigma_z$. For example, take $A=\sigma_x$. Then $A(t) = \sigma_+ e^{i\omega t} + \sigma_- e^{-i\omega t}$, and the terms $A(\boldsymbol{t}_1^m)$ have a different structure depending on the parity of $m$. For odd $m$:
\begin{equation}
A(\boldsymbol{t}_1^m) = \sigma_+ e^{i\phi(\boldsymbol{t}_1^m)} + \sigma_- e^{-i\phi(\boldsymbol{t}_1^m)}\;,
\end{equation}
while for even $m$:
\begin{align}
A(\boldsymbol{t}_1^m) &= \ket{1}\bra{1} e^{i\phi(\boldsymbol{t}_1^m)} +  \ket{0}\bra{0} e^{-i\phi(\boldsymbol{t}_1^m)}\nonumber\\
&=\one\, \cos(\phi(\boldsymbol{t}_{1}^{m})) +i\sigma_z\, \sin(\phi(\boldsymbol{t}_{1}^{m}))
\;,
\end{align}
where the phases are defined as $\phi(\boldsymbol{t}_1^m):= - \omega \sum_{j=1}^m (-1)^j t_j $.
This leads to
\begin{equation}
\Xcal{\mathbb{A}}{1}{k}{1}{n-k} = \begin{cases}
    \Xcal{\mathbb{A}_1}{1}{k}{1}{n-k}& \text{($n$ odd, $k$ odd)} \\
    \Xcal{\mathbb{A}_2}{1}{k}{1}{n-k} & \text{($n$ even, $k$ even)}  \\
    0 & \text{elsewhere}
\end{cases}\; ,
\end{equation}
with
\begin{align}
\Xcal{\mathbb{A}_1}{1}{k}{1}{n-k} =& \cos(\phi(\boldsymbol{s}_1^{n-k}))\left[\sigma_+ e^{i\phi(\boldsymbol{\tau}_1^k)} + \sigma_- e^{-i\phi(\boldsymbol{\tau}_1^k)}\right] \\
\Xcal{\mathbb{A}_2}{1}{k}{1}{n-k} =& \cos(\phi(\boldsymbol{s}_1^{n-k}))\left[\mathbb{1} + i \sin(\phi(\boldsymbol{\tau}_1^{k}))\sigma_z \right]
\end{align}

From the above it is clear that even orders give contributions on the diagonal of $K$, thus maintaining a structure proportional to $\sigma_z$, while odd orders give only contributions on off-diagonals of the effective Hamiltonian, thus changing its eigenbasis.

\bibliography{biblio}

\end{document}